\documentclass[11pt]{spie}  
\usepackage[]{graphicx}
\usepackage{glossaries}
\usepackage{hyperref}
\usepackage{subcaption}
\usepackage{mathrsfs}
\title{Wavefront Sensing in Space: Flight Demonstration II of the PICTURE Sounding Rocket Payload} 

\author{Ewan S. Douglas\supit{a}, 
Christopher B. Mendillo\supit{b},
Timothy A. Cook\supit{b}, 
Kerri L. Cahoy\supit{a,c},
Supriya Chakrabarti\supit{b}
\skiplinehalf
\supit{a}Department of Aeronautics and Astronautics, Massachusetts Institute of Technology, Cambridge, MA, USA 02139; \\
\supit{b}Lowell Center for Space Science and Technology, UMASS Lowell, Lowell, MA, USA 01854;\\
\supit{c}Department of Earth, Atmospheric, and Planetary Science, Massachusetts Institute of Technology, Cambridge, MA, USA 02139; \\
}

\authorinfo{Further author information: (Send correspondence to E.S.D.)\\E.S.D.: E-mail: douglase@bu.edu}

\usepackage{lineno}
\newacronym{AU}{AU}{Astronomical Unit [1.5e11 m]}  
\newacronym{pc}{pc}{ parsec }
\newacronym{mas}{mas}{ milliarcsecond }
\newacronym{nm}{nm}{ Nanometer  }
\newacronym{CTE}{CTE}{Coefficient of Thermal Expansion}

\newacronym{smc}{SMC}{Small Magellanic Cloud}
\newacronym{lmc}{LMC}{Large Magellanic Cloud}
\newacronym{ism}{ISM}{interstellar medium}
\newacronym{mw}{MW}{Milky Way}
\newacronym{epseri}{$\epsilon$ Eri}{Epsilon Eridani}

\newacronym{CFR}{CFR}{Complete Frequency Redistribution}

\newacronym{nasa}{NASA}{National Aeronautics and Space Agency}
\newacronym{esa}{ESA}{European Space Agency}
\newacronym{omi}{OMI}{\textit{Optical Mechanics Inc.}}
\newacronym{gsfc}{GSFC}{\gls{nasa} Goddard Space Flight Center}
\newacronym{stsci}{STScI}{Space Telescope Science Institute}
\newacronym{nsroc}{NSROC}{\gls{nasa} Sounding Rocket Operations Contract}
\newacronym{wff}{WFF}{\gls{nasa} Wallops Flight Facility}
\newacronym{wsmr}{WSMR}{White Sands Missile Range}

\newacronym{irac}{IRAC}{Infrared Array Camera}
\newacronym[plural=CCDs, firstplural=charge-coupled devices (CCDs)]{ccd}{CCD}{charge-coupled device}
\newacronym{DM}{DM}{Deformable Mirror}
\newacronym{MCP}{MCP}{ Microchannel Plate }
\newacronym{ipc}{IPC}{Image Proportional Counter}
\newacronym{cots}{COTS}{Commercial Off-The-Shelf}
\newacronym{ISR}{ISR}{Incoherent Scatter Radar }
\newacronym{atcamera}{AT}{Angle Tracker}
\newacronym{MEMS}{MEMS}{microelectromechanical systems}
\newacronym{QE}{QE}{quantum efficiency}
\newacronym{RTD}{RTD}{Resistance Temperature Detector}
\newacronym{PID}{PID}{Proportional-Integral-Derivative}
\newacronym{TRL}{TRL}{Technology Readiness Level}

\newacronym{FOV}{FOV}{field-of-view}
\newacronym{NIR}{NIR}{near-infrared}
\newacronym{PV}{PV}{Peak-to-Valley}
\newacronym{MRF}{MRF}{Magnetorheological finishing}
\newacronym{AO}{AO}{Adaptive Optics}
\newacronym{TTP}{T/T+piston}{tip, tilt, and piston}
\newacronym{FPS}{FPS}{fine pointing system}

\newacronym{acs}{ACS}{Attitude Control System}
\newacronym{orsa}{ORSA}{Ogive Recovery System Assembly}
\newacronym{gse}{GSE}{Ground Station Equipment}
\newacronym{FSM}{FSM}{Fast Steering Mirror}

\newacronym{WFS}{WFS}{wavefront sensor}
\newacronym{LSI}{LSI}{Lateral Shearing Interferometer}
\newacronym{VVC}{VVC}{Vector Vortex Coronagraph}
\newacronym{VNC}{VNC}{Visible Nulling Coronagraph}
\newacronym{CGI}{CGI}{Coronagraph Instrument}
\newacronym{IWA}{IWA}{Inner Working Angle}
\newacronym{OWA}{OWA}{Outer Working Angle}
\newacronym{NPZT}{N-PZT}{Nuller piezoelectric transducer}
\newacronym{OPD}{OPD}{Optical Path Difference}
\newacronym{WFCS}{WFCS}{Wavefront Control System}
\newacronym{SPC}{SPC}{Shaped Pupil Coronagraph}

\newacronym{HST}{HST}{ Hubble Space Telescope}
\newacronym{GPS}{GPS}{Global Positioning System}
\newacronym{ISS}{ISS}{International Space Station}
\newacronym[description=Advanced CCD Imaging Spectrometer]{acis}{ACIS}{Advanced \gls{ccd} Imaging Spectrometer}
\newacronym{stis}{STIS}{\textit{Space Telescope Imaging Spectrograph}}
\newacronym{mcp}{MCP}{Microchannel Plate}
\newacronym{jwst}{JWST}{$\textit{James Webb Space Telescope}$}
\newacronym{fuse}{FUSE}{$\textit{FUSE}$}
\newacronym{galex}{GALEX}{$\textit{Galaxy Evolution Explorer}$}
\newacronym{spitzer}{Spitzer}{$\textit{Spitzer Space Telescope}$}
\newacronym{mips}{MIPS}{Multiband Imaging Photometer for \gls{spitzer}}
\newacronym{gissmo}{GISSMO}{Gas Ionization Solar Spectral Monitor}
\newacronym{iue}{IUE}{International Ultraviolet Explorer}
\newacronym{spinr}{SPINR}{$\textit{Spectrograph for Photometric Imaging with Numeric Reconstruction}$}
\newacronym{imager}{IMAGER}{$\textit{Interstellar Medium Absorption Gradient Experiment Rocket}$}
\newacronym{TPF-C}{TPF-C}{Terrestrial Planet Finder Coronagraph}
\newacronym{RAIDS}{RAIDS}{Atmospheric and Ionospheric Detection System }
\newacronym{mama}{MAMA}{Multi-Anode Microchannel Array}
\newacronym{ATLAST}{ATLAST}{Advanced Technology Large Aperture Space Telescope}
\newacronym{PICTURE}{PICTURE}{Planet Imaging Concept Testbed Using a Rocket Experiment}
\newacronym{LITES}{LITES}{Limb-imaging Ionospheric and Thermospheric
Extreme-ultraviolet Spectrograph}
\newacronym{LBT}{LBT}{Large Binocular Telescope}
\newacronym{LBTI}{LBTI}{Large Binocular Telescope Interferometer}
\newacronym{KIN}{KIN}{Keck Interferometer Nuller}
\newacronym{SHARPI}{SHARPI}{Solar High-Angular Resolution Photometric Imager}
\newacronym{IRAS}{IRAS}{Infrared Astronomical Satellite}
\newacronym{HARPS}{HARPS}{High Accuracy Radial velocity Planetary}
\newacronym{hstSTIS}{STIS}{Space Telescope Imaging Spectrograph}
\newacronym{spitzerIRAC}{IRAC}{Infrared Array Camera}
\newacronym{spitzerMIPS}{MIPS}{Multiband Imaging Photometer for Spitzer}
\newacronym{spitzerIRS}{IRS}{Infrared Spectrograph}
\newacronym{CHARA}{CHARA}{Center for High Angular Resolution Astronomy}
\newacronym{wfirst-afta}{WFIRST-AFTA}{Wide-Field InfrarRed Survey
Telescope-Astrophysics Focused Telescope Assets}
\newacronym{GPI}{GPI}{Gemini Planet Imager}
\newacronym{WFIRST}{WFIRST}{Wide-Field InfrarRed Survey Telescope}

\newacronym{AURIC}{AURIC}{The Atmospheric Ultraviolet Radiance Integrated Code} 
\newacronym{FFT}{FFT}{Fast Fourier Transform  }
\newacronym{MODTRAN}{MODTRAN   }{ MODerate resolution atmospheric TRANsmission }
\newacronym{idl}{IDL}{$\textit {Interactive Data Language}$}
\newacronym[sort=NED,description=NASA/IPAC Extragalactic Database]{ned}{NED}{\gls{nasa}/\gls{ipac} Extragalactic Database}
\newacronym{iraf}{IRAF}{Image Reduction and Analysis Facility}
\newacronym{wcs}{WCS}{World Coordinate System}
\newacronym{pegase}{PEGASE}{$\textit{Projet d'Etude des GAlaxies par Synthese Evolutive}$}
\newacronym{dirty}{DIRTY}{$\textit{DustI Radiative Transfer, Yeah!}$}
\newacronym{MGHPCC}{MGHPCC}{Massachusetts Green High Performance
Computing Center}

\newacronym{MSIS}{MSIS}{Mass Spectrometer Incoherent Scatter Radar}
\newacronym{nmf2}{$N_m$}{F2-Region Peak density}
\newacronym{hmf2}{$h_m$}{F2-Region Peak height}
\newacronym{H}{$H$}{F2-Region Scale Height}

\newacronym{isr}{ISR}{Incoherent Scatter Radar}
\newacronym[description=TLA Within Another Acronym]{twaa}{TWAA}{\gls{tla} Within Another Acronym}
\newacronym[plural=SNe, firstplural=Supernovae (SNe)]{sn}{SN}{Supernova}
\newacronym{EUV}{EUV}{Extreme-Ultraviolet }
\newacronym{EUVS}{EUVS}{\gls{EUV} Spectrograph}
\newacronym{F2}{F2}{Ionospheric Chapman F Layer }
\newacronym{F10.7}{F10.7}{ 10.7 cm radio flux [10$^{-22}$ W m$^{-2}$ Hz$^{-1}$]  }
\newacronym{FUV}{FUV}{ Far-Ultraviolet }
\newacronym{IR}{IR}{Infrared}
\newacronym{MUV}{MUV}{Mid-Ultraviolet }
\newacronym{NUV}{NUV}{Near-Ultraviolet }
\newacronym{O$^+$}{O$^+$}{Singly Ionized Oxygen  Atom }
\newacronym{OI}{OI}{Neutral Atomic Oxygen Spectroscopic State }
\newacronym{OII}{OII}{Singly Ionized Atomic Oxygen Spectroscopic State }
\newacronym{PSF}{PSF}{Point Spread Function}
\newacronym{$R_E$}{$R_E$}{ Earth Radii [$\approx$ 6400 km]  }
\newacronym{RV}{RV}{Radial Velocity}
\newacronym{UV}{UV}{Ultraviolet }
\newacronym{WFE}{WFE}{Wavefront Error}
\newacronym{sed}{SED}{Spectral Energy Distribution}
\newacronym{nir}{NIR}{near-infrared}
\newacronym{mir}{MIR}{mid-infrared}
\newacronym{ir}{IR}{infrared}
\newacronym{uv}{UV}{ultraviolet}
\newacronym[plural=PAHs, firstplural=Polycyclic Aromatic Hydrocarbons (PAHs)]{pah}{PAH}{Polycyclic Aromatic Hydrocarbon}
\newacronym{obsid}{OBSID}{Observation Identification}
\newacronym{SZA}{SZA}{Solar Zenith Angle}
\newacronym{TLE}{TLE}{Two Line Element set}
\newacronym{DOF}{DOF}{degrees-of-freedom}
\newacronym{PZT}{PZT}{lead zirconate titanate}

\newacronym{PCA}{PCA}{Principal Component Analysis}
\newacronym{fwhm}{FWHM}{Full-Width-Half Maximum}
\newacronym{RMS}{RMS}{root mean squared}
\newacronym{RMSE}{RMSE}{root mean squared error}
\newacronym{MCMC}{MCMC}{Marcov chain Monte Carlo}
\newacronym{DIT}{DIT}{Discrete Inverse Theory}
\newacronym{SNR}{SNR}{signal-to-noise ratio}
\newacronym{PSD}{PSD}{Power Spectral Density}

\begin{document} 
  \maketitle 

\begin{abstract}
A NASA sounding rocket for high-contrast imaging with a visible nulling coronagraph, the \gls{PICTURE} payload,  has made two suborbital attempts to observe the warm dust disk inferred around Epsilon Eridani.
The first flight in 2011 demonstrated a 5 milliarcsecond fine pointing system in space.
The reduced flight data  from the second launch, on 25 November 2015, presented herein, demonstrate active sensing of wavefront phase in space. 
Despite several anomalies in flight, post-facto reduction phase stepping interferometer data provides insight into the wavefront sensing precision and the system stability for a portion of the pupil. 
These measurements show the actuation of a 32$\times$32-actuator microelectromechanical system deformable mirror. 
The wavefront sensor reached a median precision of  1.4 nanometers per pixel, with 95\% of samples between 0.8 and 12.0 nanometers per pixel. 
The median system stability, including telescope and coronagraph wavefront errors other than tip, tilt, and piston, was 3.6 nanometers per pixel, with 95\% of samples between 1.2 and 23.7 nanometers per pixel. 

\end{abstract}


\section{INTRODUCTION}
\label{sec:intro}  
Resolving reflected  light from planets in distant star systems analogous to our own  solar system requires overcoming    exoplanet-host star flux ratios  between 10$^{-9}$ and  10$^{-11}$. 
Imaging at these extreme contrast ratios\cite{turnbull_spectrum_2006,cahoy_exoplanet_2010}  will be enabled by  coronagraphs on the next generation of space telescopes. 
Coronagraphs block light from the host star and  transmit light from dim companions to a detector.
An internal coronagraph suppresses starlight between the primary mirror and the detector image plane of a telescope.
Coronagraphy requires wavefront stability, and with stellar leakage typically proportional to the square of the wavefront error dynamic and static aberrations in wavefront phase must be controlled to nanometer levels to detect self-luminous giant planets and debris disks\cite{traub_direct_2010}.
Detection of terrestrial planets requires picometer level control\cite{guyon_theoretical_2006,trauger_laboratory_2007}.
 
 The suborbital \gls{PICTURE}\cite{shao_nulling_2006,samuele_experimental_2007,rao_path_2008,mendillo_flight_2012,mendillo_picture:_2012,douglas_end--end_2015,chakrabarti_planet_2016} sounding rocket payload employs an internal \gls{VNC}, 
 behind a half-meter telescope, intended to perform high-contrast observations of exozodiacal dust in space while demonstrating internal coronagraphy along with the requisite wavefront sensing and control.

\subsection{Objectives}
 The \gls{PICTURE} observational objective was to measure the scattered light from exozodiacal dust around \gls{epseri}  at visible wavelengths (600 nm to 750 nm).
\gls{epseri} has a large infrared excess at 20 $\mu$m which has been attributed to a dusty exozodiacal debris disk\cite{di_folco_near-infrared_2007}. 
With an expected integrated brightness of  approximately 2$\times10^{-4}L_\star$, this dust may be arranged in a thin ring\cite{backman_epsilon_2009,su_inner_2017}.
 The \gls{PICTURE} missions planned to test for emission at separations from 2 AU to 20 AU, constraining  scattered light brightness and morphology and advancing our knowledge of the dust composition around sun-like stars. 
The predicted  ring of emission would provide a relatively bright target for high-contrast imaging. 
The instrument development carried out to achieve these objectives  matures coronagraphy and wavefront sensing technology\cite{chakrabarti_planet_2016}. 
 
\subsection{Wavefront Control}
Deviations from uniformity in both phase and amplitude contribute to coronagraph leakage\cite{serabyn_nulling_2000,traub_direct_2010}; however, the \gls{PICTURE} \gls{VNC} only corrects for, more significant, phase errors.
Thus, for brevity, in this work \gls{WFE} will refer solely to phase errors.
Space-based telescopes are unaffected by atmospheric turbulence, enabling diffraction limited imaging even at visible wavelengths.
In space, time-varying \gls{WFE}  primarily arises from the coupling of mechanical and thermomechanical perturbations of the spacecraft structure to the alignment and surfaces of optical components.

The ongoing Gaia mission employs a passive Shack-Hartmann wavefront sensor\cite{vosteen_wavefront_2009} and a wide variety of new wavefront sensing techniques
are being developed for future space telescopes\cite{feinberg_trl-6_2007,greenbaum_-focus_2016,shi_low_2016}. 
A simplified optical layout in Fig. \ref{fig:simple_ifo} shows how the \gls{PICTURE} \gls{VNC} interferes the white-light fringes of two offset (or ``sheared") copies of the input telescope pupil in a Mach-Zender instrument with a $\pi$ phase offset between the two pupils. 
Active wavefront phase sensing is implemented in the  \gls{PICTURE} \gls{VNC} using phase stepping interferometry\cite{wyant_use_1975}, where the intensity is measured at multiple points across the fringe pattern by changing the relative path length before beginning science observations.
Basics of the \gls{VNC} wavefront sensing are further described in the Appendix and details of the coronagraph implementation and optical design are given elsewhere\cite{rao_path_2008,mendillo_picture:_2012,chakrabarti_planet_2016}.

\subsection{Deformable Mirrors}
\gls{DM}s enable precision wavefront control by dynamically minimizing the \gls{WFE} across a range of spatial frequencies. 
The \gls{PICTURE} \gls{VNC} employs a \gls{DM} to minimize the path  difference between the between the two arms of the interferometer, enabling the nulling of starlight even in the presence of wavefront error between the sheared pupil sub-apertures. 
Numerous \gls{DM} technologies exist or have been proposed, including piezoelectric\cite{ealey_xinetics_1994}, thermoelectric coolers\cite{huang_experimental_2015}, ferrofluid \cite{lemmer_mathematical_2016}, and \gls{MEMS}\cite{bifano_microelectromechanical_1999}.
Compact size, high-actuator count, low-power consumption, and extensive use in ground-based adaptive optics\cite{morzinski_mems_2012}  make \gls{MEMS} \gls{DM}s particularly desirable for space-based applications.
\gls{MEMS} \gls{DM}s typically rely on voltages up to 250 V to electrostatically displace a membrane on sub-angstrom to micron scales\cite{bifano_microelectromechanical_1999}. 
There are few references to \gls{MEMS} optical device operation in space.
Yoo et al.\cite{yoo_mems_2009} found a non-deformable (on-off only) \gls{MEMS} micromirror device maintained  functionality after undergoing launch and was successfully operated on the \gls{ISS}.
A magnetically actuated \gls{MEMS} microshutter array has been flown on the Far-ultraviolet Off Rowland-circle Telescope for Imaging and Spectroscopy (FORTIS) sounding rocket\cite{fleming_calibration_2013}.
Finite element modeling has been used to predict the survival of \gls{MEMS} \gls{DM}s subjected to launch\cite{aguayo_fem_2014}, but survival in a harsh launch environment has not been found previously demonstrated in the literature.  
                                    
The \gls{PICTURE} \gls{DM} is a Boston Micromachines 32 $\times$ 32 square Kilo-DM with a 1.5 $\mu$m stroke, a continuous gold-coated silicon membrane, a 340 $\mu$m actuator pitch, and custom drive electronics\cite{rao_path_2008}. 
Actuator drive voltages are limited to $\leq$ 150 V to prevent actuator snap down\cite{morzinski_mems_2012}.
To avoid chromatic optical path mismatch between the two \gls{VNC} arms, the \gls{DM} lacks the protective window commonly included to prevent  mirror surface  contamination.

\subsection{Sounding Rocket Flights}

Flight I of the \gls{PICTURE} payload launched\cite{mendillo_flight_2012,mendillo_picture:_2012,hicks_nulling_2012,mendillo_scattering_2013} aboard \gls{nasa} sounding rocket 36.225 UG (a Black Brant IX University Galactic Astronomy mission) on 8 October 2011 from \gls{wsmr}. 
The  flight  suffered a telemetry failure approximately seventy seconds after launch during calibration observations of Rigel.
Unfortunately, limited \gls{WFS} data was transmitted before a telemetry relay failed\cite{yuhas_sounding_2012} and no interference fringes were observed, preventing wavefront sensing or nulling. 
A second mission,  renamed PICTURE-B (Planet Imaging Coronagraphic Technology Using a Reconfigurable Experimental Base) refurbished and relaunched the same payload with minor modifications in the fall of 2015.  
The results of this nearly identical mission, which had the same science goals as the original flight, are discussed in this paper.

The flight data presented herein was collected during flight of the refurbished payload, \gls{PICTURE}-B (NASA 36.293 UG), launched from \gls{wsmr} at 9:17 p.m. MST November 24th (25 November  2015 0417 UT).
The payload design, concept of operations, and the Flight I telemetry failure were described previously by Mendillo et al.\cite{mendillo_picture:_2012}, and refurbished telescope and nulling coronagraph integration and laboratory performance were summarized by Chakrabarti et al\cite{chakrabarti_planet_2016}. 
 Fortunately, the Flight II (36.293) telemetry system performed as designed, and data were redundantly stored onboard, providing far more insight into instrument performance in flight.
In this work we focus on the operation of the \gls{WFS} and \gls{DM} in space during Flight II.
  Section \ref{sec:refurb} briefly summarizes the refurbishment of the payload.
  Section \ref{sec:observations} describes the flight observation sequence and the anomalous \gls{WFS} measurements recorded in flight.
     Section \ref{sec:methods}  details the post-processing methods used to interpret the flight data.
    Section \ref{sec:results} presents the estimated  precision of these measurements. 
  Finally,  Section  \ref{sec:conclusion} remarks on conclusions and future directions.

\section{Refurbishment}\label{sec:refurb}
\subsection{Deformable Mirror}
The cabling to the \gls{DM} was damaged during the assembly of the payload for Flight I.
Thus, a new polyimide flex cable assembly was manufactured and installed along with a replacement Boston Micromachines Corporation  Kilo-DM (S.N. 11W310\#002).
MEMS DM actuator yield has improved greatly over time: early Kilo-DM models had yield as low as 96.9\%\cite{evans_demonstrating_2006}, whereas this \gls{DM} has two inactive actuators (a 99.8\% yield). 
Fortunately,  both inactive actuators were positioned behind the Lyot mask, which blocks un-interfered light from behind the sheared secondary obscurations, allowing phase control across the entire output pupil.
To best match reflectivity between the interferometer arms, a new \gls{NPZT} mirror,  coated in the same chamber as the replacement \gls{DM}, was also installed. 
Located in the \gls{VNC} arm opposite the \gls{DM}, the \gls{NPZT} mirror is mounted on a Physik Instrumente  S-316 piezoelectric stage, driven by custom electronics,  which corrects \gls{TTP} errors between the two interferometer arms with maximum piston range of approximately 8.5 $\mu$m.
The two optics were aligned in the laboratory such that the \gls{NPZT} mirror flattened the wavefront error in the \gls{VNC} at room temperature in the middle of the piezoelectric stage's range. 

\subsection{Preflight Testing}\label{sec:preflight_testing}
The \gls{VNC} was tested post-refurbishment, without the telescope, using a simulated point source and a retro-reflecting mirror.
The contrast was found to be comparable to previous tests with central star leakage of approximately $10^{-3}$\cite{rao_path_2008,chakrabarti_planet_2016}.
The \gls{VNC} residual wavefront error at \gls{DM} controllable spatial frequencies was corrected to an \gls{RMSE} phase of 5.7 nm$\pm$2.6 nm\cite{douglas_advancing_2016}.
The fully assembled payload was tested end-to-end on a vibration suppressing optical table and nulling was observed \cite{chakrabarti_planet_2016}.

The integrated payload was shake-tested at \gls{wsmr} to Vehicle Level Two random thrust vibration\cite[Table 6.3.4-1]{noauthor_sounding_2015}. 
The sounding rocket payload assembly provides some damping and the integrated acceleration of the nuller assembly during the random vibration was 10.4 g \gls{RMS} versus the  12.7 g \gls{RMS} input. 

\section{Observations}\label{sec:observations}

\subsection{Target Acquisition}
Both Flight I and II used  Rigel  ($\beta$ Orionis, $m_v$=0.13) as the initial calibration star. 
Unfortunately, neither flight successfully advanced from Rigel to \gls{epseri}. 
During both missions, a manual uplink successfully provided the pointing correction to the \gls{acs}, placing Rigel on the angle tracker camera near the nominal t+105 second observing start time.
 Additional  uplinks commanded centering of the star on the angle tracker to the \gls{acs} accuracy of approximately 1 arcsecond.
Once Rigel was centered, the \gls{FPS}  control loop locked, providing pointing precision of approximately five milliarcseconds (1$\sigma$), consistent with Flight I\cite{mendillo_flight_2012}. 

After acquisition and \gls{FPS} lock, three attitude control system maneuvers were planned:
1) Nuller alignment and 10 seconds of speckle observations on Rigel,
2) Slew to \gls{epseri} and observe the circumstellar environment, 
3)  Roll payload 90 degrees during the \gls{epseri} observation to characterize speckles.
Flight II did not complete nuller alignment, but a number of on-star observations were recorded.

\subsection{Data Products}\label{sec:data_products}
Two cameras observed the \gls{VNC} output after the Lyot stop, which transmitted only regions where the two sheared interferometer arms overlap, the science camera to image the sky and the \gls{WFS} camera to image the interference fringes in the pupil plane. 
Cut-off filters limited the observed bandwidth of both cameras to between 600 nm and 750 nm.
Due to lack of nulling, the science camera saturated throughout the flight and will not be considered further.
The  \gls{WFS} and science cameras were \gls{ccd} detectors developed for the Astro-E2 X-ray Imaging Spectrometer\cite{bautz_progress_2004}. 
These 1024$\times$1024 pixel MIT Lincoln Laboratory model CCID41 detectors were cryogenically cooled to -70$^\circ$C for a dark noise of approximately 1 e$^-$/s/pixel. 
In order to allow short exposure times, only small subregions were read-out from each camera. 
The \gls{WFS} readout area was 76 pixel $\times$ 76 pixel.
The integration time of each \gls{WFS} frame time was 0.23 seconds.
The laboratory measured read noise rate was 2.3 e$^-$ per pixel per exposure. 
 
\subsection{Wavefront Sensing Data}\label{subsec:example_WFS}

During Flight II the \gls{WFCS} advanced to the \gls{VNC} phase-up or ``coarse mode" [detailed in Fig 9]\cite{mendillo_picture:_2012} stage of wavefront sensing and low-order correction.
This included locating the white-light fringe packet; applying a predetermined set of voltages to the \gls{DM}, thereby removing the stress-induced surface concavity from fabrication\cite{bifano_micromachined_2000}; and low-order \gls{WFE} correction of between the interferometer arms with the \gls{NPZT} mirror in \gls{TTP}.
In the planned flight sequence, phase-up was followed by fine correction of higher spatial frequencies with the \gls{DM} and finally a transition to \textit{nulling mode} with the \gls{NPZT} shifted to the dark fringe for high-contrast science observation over the remainder of the flight.

Flight II did not reach the fine-mode correction or nulling modes because the wavefront could not be flattened, likely due to shift in the \gls{DM} mount, as discussed in Sec. \ref{sec:discussion}. 
Fringe visibility was used as the in-flight metric of coherence and is calculated from the maximum and minimum intensity in a pixel, $I$: $\mathscr{V}\equiv ({I_{max}-I_{min}})/({I_{max}+I_{min}}).$
Black dots in Fig. \ref{fig:visibility_vs_time} show this median $\mathscr{V}$ versus time at the \gls{WFS}' approximately 1 sec cadence.
The closed-loop correction of wavefront mismatch between the arms with the \gls{DM} was not initiated because  the fringe packet was outside the range of the \gls{NPZT} to correct \gls{TTP} errors, keeping $\mathscr{V}$  below the threshold to advance modes.  
The magnitude of this misalignment is estimated in Section \ref{sec:discussion}.
The phase-up flattening mode required accurate measurements of the \gls{WFE}.
Thus, the mission  returned  measurements of phase error and \gls{WFS} stability; the remainder of this analysis will focus solely on this \gls{WFS} camera data.

The example sets of interferometric \gls{WFS} images shown in Fig. \ref{fig:ABCDflight} typify the range of raw measurements recorded over the course of the flight.
While the wavefront was not flattened, these fringes are visible across the pupil plane images recorded by the \gls{WFS} camera and provide insight into the alignment of the \gls{VNC} and the state of the \gls{DM}.
Each row  of Fig. \ref{fig:ABCDflight} is an example set of fringe measurements from different times in the flight. 
Each set of fringe measurements corresponds to a wavefront sensor image at four unique \gls{OPD} points, separated by $\pi/2$ radians, which together are used to reconstruct the \gls{WFE} at a point in the flight (see  illustration in Fig. \ref{fig:simple_ifo} and derivation in Appendix).
The left four columns illustrate the background subtracted raw intensity. (Background noise levels were calculated from the median of \gls{WFS} exposures recorded in-flight  before and after the Rigel observation.)  
When the system was first powered on the \gls{WFS} fringe $\mathscr{V}$ was low, as seen in the top row of Fig. \ref{fig:ABCD26_28}. 
The \gls{NPZT} was functioning normally, stepping $\pi/2$ steps, but far from the center of the coherent fringe packet, as the fringe flattening \gls{TTP} correction had not yet been applied.
The sequence of high-$\mathscr{V}$ measurements, as seen in the bottom row of Fig.  \ref{fig:ABCD26_28}, were recorded at the beginning of the flight when the flight software was attempting to flatten the \gls{OPD} across the pupil (the analysis period, shaded in Fig. \ref{fig:visibility_vs_time}). 
The bottom row of Fig.  \ref{fig:ABCD26_28}  is representative of this series of phase-up  measurements   which will be used for the bulk of the following analysis. 
The map of per-pixel $\mathscr{V}$ in the rightmost column of Fig.   \ref{fig:ABCD26_28} is higher on the left side of of the \gls{WFS} image but decreases towards the lower right because the \gls{NPZT} actuator in that corner is out of range and ``railed'' at the maximum displacement, indicating the the central white light fringe was just beyond the range of the stage.
The geometry of this anomalous behavior is shown schematically in the inset of Fig. \ref{fig:simple_ifo}.

A high visibility, $\mathscr{V}>0.9$ across the pupil was expected; however, the median flight visibilities never exceeded 0.8 due to incomplete stepping of the \gls{NPZT} mirror, as will be described in  Sec. \ref{subsec:example_WFS}.
Before the telescope shutter closed for reentry, several unsuccessful attempts were made to repoint the payload via human-in-the-loop command uplinks from the ground to bring the \gls{NPZT} into piston range. 
 This caused low $\mathscr{V}$ during each pointing maneuver and  re-acquisition of Rigel by the \gls{FPS}. 
  Closed-loop \gls{FPS} control was lost during repointing, as seen in the ``lock'' status of the \gls{FPS} angle tracker.
  This status is plotted as a horizontal line in Fig. \ref{fig:visibility_vs_time}.
  The futility of these attempts is seen in the increased variability of the $\mathscr{V}$ as a function of time, as telescope focus deteriorated and the coherent fringe packet drifted further out of the \gls{NPZT} range.
The focus degraded due to thermal gradients across the telescope optical bench (see Douglas\cite{douglas_advancing_2016} for details on the thermal environment and the evolution of the \gls{PSF} as measured by the \gls{FPS} camera).

\section{Methods}\label{sec:methods}
The \gls{PICTURE} design leverages the interferometric nature of a nulling coronagraph to directly measure wavefront error by imaging the pupil at the science output of the \gls{VNC}.
Before nulling, the wavefront into the system must first  be measured and corrected to flatten the interference fringe.
This section describes issues encountered in flight, and post-facto analysis to retrieve the phase which will be used to quantify \gls{WFS} performance and show actuation of the deformable mirror surface.

The \gls{VNC} control software expected $\pi/2$ phase steps 
 to calculate the \gls{WFE}  error (Appendix Eq. \ref{eq:phi_ABCD}) to be corrected by the \gls{NPZT} and \gls{DM}\cite{mendillo_picture:_2012}.
During Flight II one of the three piezo actuators translating the \gls{NPZT} mirror was railed high for many of the ABCD measurements, while the other two actuators moved the mirror in $\pi/2$ steps, causing a varying phase shift ($\delta$) across the pupil image.

This is particularly true of the high-$\mathscr{V}$ measurements (e.g., bottom panels of Fig. \ref{fig:ABCD26_28}) where the path length between the arms was best matched, meaning the railed images are also the measurements with the most coherent interference.
In order to better estimate the uncertainties in the returned data given the anomalous \gls{NPZT} steps,  raw intensity maps are re-reduced to find the best-fit phase in each pixel.
We recover the \gls{WFE} from these maps by fitting the phase in each pixel, as described in the Appendix.

\section{Wavefront Sensor Results}\label{sec:results}
This section presents a stability analysis of the in-flight wavefront error during the period of consecutive phase-up measurements recorded after the \gls{FPS} first locked and before repointing was attempted. 
These ten ABCD image sets and  phase measurements, calculated a described in the Appendix, correspond to the best instrument focus and the highest-$\mathscr{V}$ fringes.
 The mean precision of each phase measurement, mapped in Fig. \ref{fig:lsq_4panel}(a), is the standard deviation of the phase fitting error, including photon noise, for each of the pixels across the pupil.    
This mean measurement error across the interfering pupil becomes highly uncertain (exceeding $\pi$ radians) where the phase steps become small due to lack of  relative motion of the \gls{NPZT} mirror.

In order to compare the wavefront sensor precision to the expected performance of a fully stepping wavefront sensor, we define a sufficiently stepping region and exclude the regions of the pupil where the \gls{NPZT} step-size was relatively small (below $\pi/4$). 
This sufficiently stepped region has an area of 759 \gls{WFS} pixels (35\% of the pupil area), providing a relatively large sample with which to assess the instrument stability and sensing precision.    
 The mean of the standard  fitting error across this region,  Fig. \ref{fig:lsq_4panel}(a), is less than $\lambda/2$ and relatively uniform.

To assess the wavefront sensing stability, \gls{TTP} errors are subtracted from each phase measurement by fitting a two dimensional plane to the remaining phase pixels, because the \gls{NPZT} was operating in a closed-loop correction mode and each \gls{NPZT} position varied slightly.
Due to the shearing mechanism of the \gls{VNC}, this approach also removes errors due to changes in telescope focus and astigmatism\cite{douglas_advancing_2016}.
The mean map of the \gls{TTP} subtracted phase measurements in the >$\pi/4$ step region, Fig. \ref{fig:lsq_4panel}(c), is relatively uniform with a drop off at the bottom of the map where the uncertainty is also higher.
The system stability, including both the telescope and \gls{VNC}, is measured by taking the standard deviation of the recovered phase in each pixel across the ten recorded \gls{WFS} sequences, Fig. \ref{fig:lsq_4panel}(d).


The distributions of $1\sigma$ uncertainty for both the measurement error  (dark gray, Fig. \ref{fig:wfs_sigma_hist}) and the system stability (light blue, Fig. \ref{fig:wfs_sigma_hist}), exhibit strong peaks at a few nanometers with median values of 1.4 nm/pixel and 3.6 nm/pixel, respectively.
 These distributions of \gls{WFE} per pixel  exhibits  long tails, or a few pixels with very high uncertainty.
 These pixels corresponds to the low-$\mathscr{V}$ region at the bottom edge of the pupil. 
Since the distributions are highly skewed, the range of values is well represented by calculation of the sample interval, the percent of values within a range.
Sample intervals were calculated using the  \textit{mquantiles} statistics function with default values as implemented in SciPy\cite{jones_scipy:_2001} version 0.19.
A 95\% sample interval of the wavefront stability spans from 1.2 nm/pixel  to 23.7 nm/pixel.
The wavefront sensor fitting error sample interval spans from 0.8 nm/pixel to 12.0 nm/pixel.

 
%



\subsection{Discussion}\label{sec:discussion}
The  in-flight wavefront error lacks  the stress-induced surface concavity expected for the unpowered \gls{DM}. 
This is illustrated in Fig \ref{fig:dm_off_meas},  the left panel presents an unpowered laboratory surface measurement, rescaled to correspond to wavefront error in the interferometer ($2\cos(15^\circ)$, where $15^\circ$ is the angle of incidence), and masked before subtraction of \gls{TTP} terms.
The unpowered \gls{WFE} exceeds a micron and exhibits a clear roll-off toward the edges of the square \gls{DM}.
In contrast, the  \gls{WFE} measured in flight (right panel, Fig. \ref{fig:dm_off_meas})  is nearly flat, with a few hundred nanometer downturn at the bottom of the pupil, where measurement errors are highest (Figs. \ref{fig:lsq_4panel}(b) and (d). 
This indicates that the \gls{DM} was powered on, and the measured actuators were responding appropriately, moving to the commanded default positions.

The observed fringe $\mathscr{V}$ provides a first-order estimate of the \gls{NPZT} mirror position relative to the center of the fringe packet. 
The translating mirror was originally aligned to the mid-point of the 8.5 micron range.
As discussed in the Appendix, the coherence length of the \gls{VNC} is approximately 3 $\mu$m. 
The visibility was moderate, with approximately one wave of phase tilt across the pupil  in the first measurement (Fig. \ref{fig:ABCD26_28}, top row). 
Thus, the central fringe was within a few microns of the \gls{NPZT} limit, implying a total motion of no more than half the range plus the coherence length or 7 $\mu$m  from the original alignment at the center of the \gls{NPZT} range. 
Post-flight laboratory testing found a  displacement consistent with this estimate; the optimal \gls{NPZT} mirror position has shifted several microns compared to the prelaunch alignment, implying minimal movement during re-entry and recovery. 

This shift is likely due to motion of the 6-degrees-of-freedom \gls{DM} mount. 
Whether this shift occurred due to launch forces or upon reentry and impact cannot be definitively determined since a flight shift of the \gls{NPZT} or \gls{DM} mounts could also have been due to a large temperature gradient within the payload.
However, the \gls{DM} mount temperature was stable to approximately 0.3$^\circ$ C in flight\cite{douglas_advancing_2016} and prior laboratory tests found the \gls{VNC} path length has an approximately 700 nm/$^\circ$C dependency\cite{mendillo_scattering_2013}, well within the range of the \gls{NPZT} to correct.
 Random vibration is a large contributor to optical bench instability in spacecraft\cite{edeson_dimensional_2009}, making the sounding rocket launch environment the most likely cause of a few micron displacement. 
This suggests improved mount designs, or additional active correction stages, are required for future missions with micron-scale alignment tolerances.
While the payload underwent random vibration before launch, the launch of a sounding rocket also subjects the payload to additional acoustic loads (sound pressure levels typically exceed 100 dB \cite{griffin_active_2008}), as well as continuous vertical launch acceleration and centripetal acceleration due to the 4 Hz rotation of the sounding rocket\cite{noauthor_sounding_2015}. 
Constant acceleration in particularly is difficult to replicate in testing and may have contributed inelastic deformation or slippage of the \gls{DM} mounting structure.

The stability of the \gls{WFS} measurements indicates the \gls{DM} surface and the $\delta$ step positions were relatively stable and lower than the laboratory measured stability for the \gls{VNC} alone (Sect. \ref{sec:preflight_testing}). 
Four wavefront sensor pixels sample each \gls{DM} actuator; thus, assuming the actuator errors are uncorrelated, the uncertainty in phase per actuator is half the values reported herein.
The measurement error in the sufficiently stepping region reached the expected 2 nm wavefront error floor\cite{mendillo_scattering_2013} due to photon noise for Rigel.
Including the instability, had these \gls{WFS} measurements been applied to the \gls{DM} to correct phase mismatch between the \gls{VNC} arms,   theoretical total leakage per actuator due to residual phase error\cite{serabyn_nulling_2000} would have been approximately $1\times10^{-4}L_\star$, exceeding laboratory performance.

\section{Conclusions}\label{sec:conclusion}
The \gls{PICTURE} sounding rocket program has advanced exoplanet imaging technology by translating laboratory demonstrated concepts into deployed spaceflight hardware.
The PICTURE program has previously demonstrated an \gls{FPS} that provides precision pointing, and this analysis of the second PICTURE flight shows active wavefront sensing precision at nanometer scales with a \gls{VNC}.
The higher-order \gls{WFE} was stable over the ten second analysis period, ten times faster than the control loop update rate\cite{mendillo_picture:_2012}, indicating the residual \gls{WFE} was correctable by the \gls{DM} had the loop closed. 
The difficulty in operating the \gls{VNC} in space illustrates the challenge of maintaining the alignment of an interferometer.
This suggests future work is necessary to increase the robustness of \gls{VNC}s if they are to be alternatives to more robust coronagraph designs, such as the \gls{SPC}\cite{kasdin_extrasolar_2003} or the \gls{VVC}\cite{mawet_vector_2010}. 
One example  of a structurally robust interferometer concept is the monolithic achromatic nulling interference coronagraph\cite{hicks_monolithic_2009}.
 
 The observations described herein also demonstrate the first operation and measurement of a \gls{DM} for high-contrast imaging in space in reflected light. 
Several upcoming missions will continue  progress in high-contrast imaging from space over longer durations and/or larger telescope apertures.
The  Planetary Imaging Concept Testbed Using a Recoverable Experiment - Coronagraph (PICTURE -- C) high-altitude balloon will demonstrate wavefront sensing and control of both phase and amplitude with a \gls{VVC}\cite{cook_planetary_2015,mendillo_optical_2017} over a timescale of hours with a half-meter telescope. 
The \gls{DM} Demonstration Mission CubeSat\cite{douglas_design_2017} is being built as a \gls{MEMS} \gls{DM} testbed to operate for over a year in low-Earth with an few centimeter diameter telescope.
  The  \gls{WFIRST} coronagraph instrument\cite{spergel_wide-field_2015} is planned to demonstrate wavefront sensing and control for both \gls{SPC} and hybrid-lyot internal coronagraphs behind an obscured telescope aperture during a multiyear mission.

\acknowledgments     

 The PICTURE-B team would like to thank the NASA Sounding Rocket Program Office, the Wallops Flight Facility, and the Orbital ATK  NSROC II team for their support, particularly our mission managers: Christine Chamberlain and David Jennings. 
We are also deeply indebted to everyone in the \gls{wsmr} Naval Research Rocket Support Office and NMSU Physical Science Laboratory teams for their leadership and assistance.
 
This work was supported by NASA grants NNG05WC17G, NNX11AD53G, NNX13AD50G, \\NNX15AG23G, and through graduate fellowships awarded to E.S. Douglas by the Massachusetts Space Grant Consortium.   
Computing resources were provided by two interfaces to the Massachusetts Green Computing Facility:  MIT Research Computing and the Boston University Scientific Computing Cluster. 

Special thanks to Brian A. Hicks of NASA Goddard Space Flight Facility, Benjamin F. Lane of MIT Draper Laboratory, and Shanti Rao and J. Kent Wallace of the Jet Propulsion Laboratory. 
The Boston University Scientific Instrument Facility worked tirelessly to support to integration of both PICTURE  payloads. 
 Paul Bierden, Charles Conway, and the rest of the staff of Boston Micromachines Corporation provided invaluable support to this project.  
 The staff at AOA Xinetics Northrop Grumman and John G. Daly of Vector Engineering provided essential support to the refurbishment of the flight telescope.
E.S.D. would especially like to thank Catherine Espaillat, Alan Marscher, Donald W. McCarthy, and Michael Mendillo for their valuable input.
 This research made use of community-developed core Python packages, including: Astropy\cite{the_astropy_collaboration_astropy:_2013}, Matplotlib\cite{hunter_matplotlib:_2007}, SciPy\cite{jones_scipy:_2001}, and
the IPython Interactive Computing architecture \cite{perez_ipython:_2007}.
Additional data analyses were done using IDL  (Exelis Visual Information Solutions, Boulder, Colorado).  
This research has made use of the SIMBAD database, operated at CDS, Strasbourg, France.

 \section*{Disclosures}     
The authors have no financial interests to disclose. 

\section{Appendix: Calculating the phase}\label{subsec:phasecalc}
In the  nulling coronagraph architecture\cite{bracewell_detecting_1978} (``nuller") two equal intensity beams of quasi-coherent starlight, collected by apertures separated by a baseline (interferometer ``arms'') with a relative phase shift of $\pi$ are combined to form a fringe pattern on the sky.
When recombination occurs at a beamsplitter, 
the output is divided into two paths:  the ``dark fringe" path where starlight  destructively interferes (``nulling''), and a second path where light constructively interferes, the ``bright fringe''.
Nulling coronagraphy requires coherent interference, which means the absolute path between each interferometer arm must be matched, otherwise the beams are temporally incoherent and interference fringes will not be observed. 
Within a few wavelengths of this absolute phase shift light  remains quasi-coherent, with the \gls{VNC}  bandwidth in wavelength space defining the coherence length of this interference fringe packet. 
The relative path differences between the two arms of the interferometer depend on the source angle on the sky with respect to the optical axis. 
Thus, when the fringe pattern is centered on a star, the light from exoplanets at small angular separations is partially transmitted while the starlight is nulled.

The \gls{PICTURE} \gls{VNC} design\cite{rao_path_2008} is a uni-axial Mach-Zehnder \gls{LSI} design\cite{shao_visible_2004,shao_calibration_2005,lyon_visible_2006,levine_visible_2006} with dispersion plates which allow for broadband nulling\cite{morgan_nulling_2000}. 
By splitting the input wavefront with a beamsplitter  and offsetting the two arms laterally, the \gls{LSI} design allows for interference between two sub-apertures formed from a single telescope pupil at the second beamsplitter.

 Simplifying the interference equation\cite{born_principles_1980} by assuming two beams of equal intensity ($I$) gives a relation between the phase difference, $\Delta\phi$, and the fringe intensity, $I(\Delta\phi$) between the beams:
  \begin{equation}\label{eq:intensity_phase}
  I(\Delta\phi)=2I+2I\cos(\Delta\phi)\mu.
  \end{equation}
  Here $\mu$ is the coherence between the two beams. 
    $\mu$ is near unity for measurements at the center of the interference fringe packet.
The total phase difference can be written as $\Delta\phi=\delta+\Delta\phi'$ where $\Delta\phi'$ is the \gls{WFE} and $\delta$ is a known relative phase step between separate measurements.
This allows expansion of the cosine term: $\cos(\Delta\phi)=\cos\delta\cos\Delta\phi'-\sin\delta\sin\Delta\phi'$. 
Defining three new variables allows us to simplify the relation, $a_0=2I$, $a_1=a_0\cos\Delta\phi'$, and $a_2=-a_0\sin\Delta\phi'$, such that: $I(\Delta\phi)=a_0+a_1\cos\delta+a_2\sin\delta$.

 The \gls{PICTURE} \gls{VNC} was designed to recover phase by recording \gls{WFS} intensity measurements as a sequence of four measurements separated by $\pi/2$. 
 For convenience, we rename each of these intensities: \textbf{A}$=I(\delta=0)$,  \textbf{B}$=I(\delta=\pi/2)$,  \textbf{C}$=I(\delta=\pi)$,  \textbf{D}$=I(\delta=3\pi/2)$. 
Solving the system of equations composed of the four intensity measurements and the known phase step values  permits calculation of the \gls{WFE} of each pixel in a set of \textbf{ABCD} measurements\cite{wyant_phase-shifting_2011,wyant_use_1975}:

\begin{equation}\label{eq:phi_ABCD}
\Delta\phi' = \arctan(\frac{A-C}{B-D}).
\end{equation}

Interference fringes in intensity due to $\Delta\phi'$ are visible when the path lengths are matched to within the coherence length of the fringe packet.  
(For the 150 nm bandpass, the coherence length at 675 nm is approximately 3 $\mu$m.)

To compensate for the uneven shifting of the \gls{NPZT} mirror,  an alternative approach to measuring phase was applied.
For varying values of $\delta$, the phase error ($\Delta\phi'$) can be recovered by least-squares fitting of the intensity ($I(\Delta\phi)$) versus phase step-size ($\delta$). 
$\delta$ values for each \gls{WFS} pixel were calculated from commanded \gls{NPZT} positions using a laboratory calibrated transformation matrix of \gls{NPZT} data values to the \gls{TTP} values in units of distance.
The resulting step map is shown in Fig. \ref{fig:RailedStepMap}. 
Allowing for variation in coherence, we again expand Eq. \ref{eq:intensity_phase} and fit  a model of three unknowns:
\begin{equation}
I(\Delta\phi)=a_0+(a_0\cos\Delta\phi'\cos\delta-a_0\sin\Delta\phi'\sin\delta)\mu.
\end{equation}

To constrain the problem, bounds were set requiring a coherence between $1\times10^{-9}$ and unity and a phase shift between $0$ and $2\pi$. 
This least squares bound-constrained minimization was solved using the subspace trust region interior reflective algorithm\cite{branch_subspace_1999} implemented in SciPy 0.19\cite{jones_scipy:_2001}. 
Least-squares fitting of each pixel was repeated on the four images of a railed measurement with varying values of $\delta$. 
The left panel of Fig. \ref{fig:lsq_phi_maps} shows the resulting phase map.
These phase measurements wrap about $2\pi$ radians and were unwrapped in order of pixel reliability in a noncontiguous fashion via the Herr\'{a}ez\cite{herraez_fast_2002} method, middle panel of Fig. \ref{fig:lsq_phi_maps}.
The rightmost panel of Fig. \ref{fig:lsq_phi_maps}  shows the corresponding fitting error, including the photon noise, calculated by taking the diagonal of the covariance matrix. 
The phase error measurement rapidly deteriorates once the phase step across the pupil drops below $\pi/4$ due to the railed actuator.

\bibliography{../../../MyLibrary}   
\bibliographystyle{spiebib}   

\newpage

\begin{figure}[tp]
   \centering  
       			\includegraphics[width=0.5\textwidth]{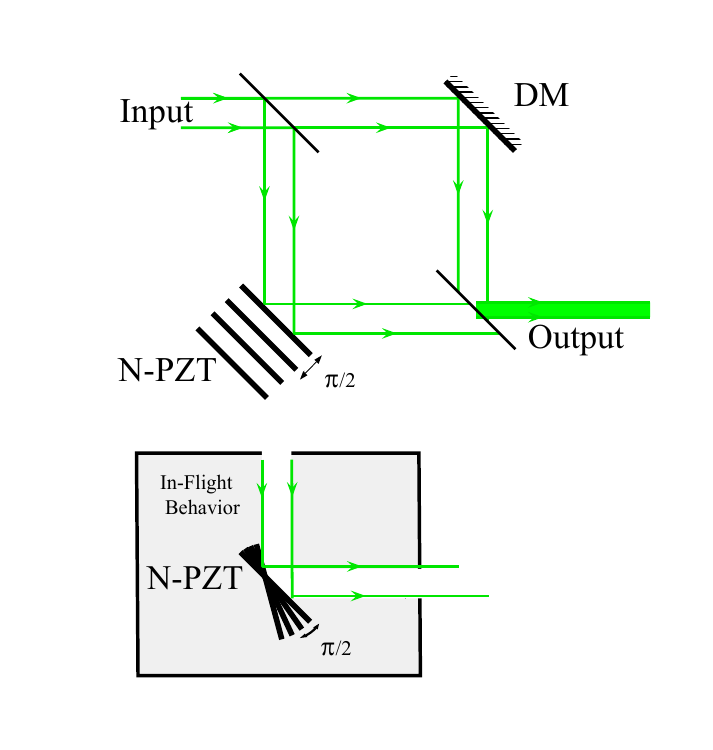}
		\caption{Simplified \gls{PICTURE} \gls{VNC}  layout and \gls{WFS} operation.
		  In normal operation (top), the mirror mounted on the \glsname{NPZT} stage steps though four positions, separated by $\pi/2$ radians, allowing reconstruction of the phase difference between the two interferometer arms at the output. 
		The inset (bottom) shows the \glsname{NPZT} mirror tilted about one of its three actuators while searching for the central white-light fringe, which had moved out of range of the stage.  
		The gray inset shows how one of the three \glsname{NPZT} actuators was at its maximum extent and thus the mirror performed incomplete steps.
		Pupil shearing elements and dispersive phase plates  omitted for clarity. Not to scale.}\label{fig:simple_ifo}
\end{figure}

\begin{figure}[tp]
   \centering  
   \includegraphics[width=0.5\textwidth]{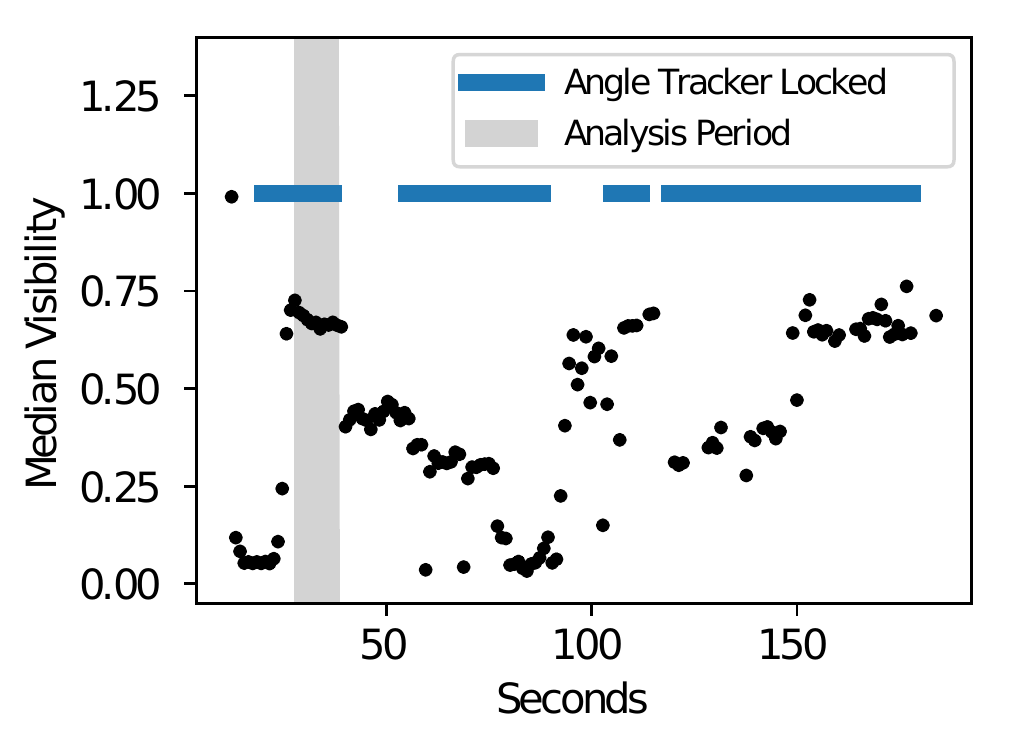}
    \caption{Visibility versus frame number for flight observations of Rigel.
    Each point represents the median visibility across all illuminated pixels of a  \gls{WFS} measurement. 
    The  horizontal line at unity represents the locked state of the \gls{FPS},  breaks in the line occur at times when the telescope was repointed causing the visibility to change. 
The shaded  gray region represents the period used to calculate \gls{WFS} performance. 
In the later \gls{FPS} locked periods, the telescope focus had deteriorated, which decreased the \gls{FPS} stability and consequently increased the variability in the visibility. 
}
      \label{fig:visibility_vs_time}
\end{figure}

\begin{figure}[tp!]
\begin{centering}
       			\includegraphics[width=\textwidth]{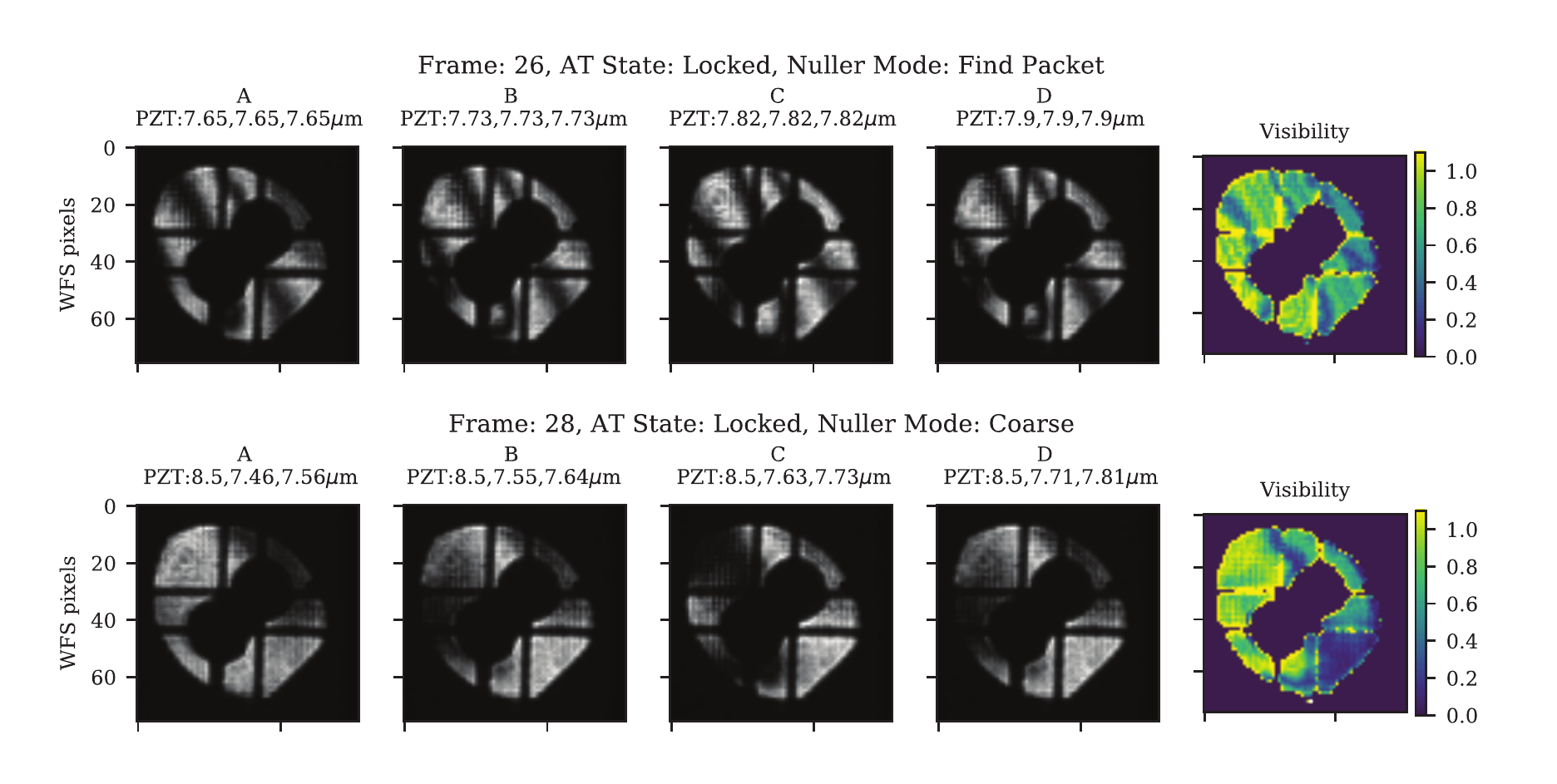}	
          \caption{ Examples of raw  wavefront sensor intensity measurements during Flight II of the star Rigel.
    The first four grayscale columns correspond to the four \gls{NPZT} positions (\textbf{A}, \textbf{B}, \textbf{C}, and \textbf{D}) each shifted by a quarter-wave. 
    The position of each of the three \gls{NPZT} actuators is shown in microns in the title of each image, these motions correspond to half of the relative wavefront shift, $\delta$.
    The first actuator listed is unchanging and railed at 8.5 $\mu$m for each image in the top and bottom rows.
    The far right column shows the $\mathscr{V}$ of each row of measurements.
    			Top: {Example measurement with complete phase steps and the \gls{NPZT} mirror in the initial flat position, leaving a large tilt relative to the input beam. 
				Since each step is complete, the fall-off in $\mathscr{V}$ across the pupil is due to the finite extent of the fringe packet. }
				Bottom: {The first measurement recorded in coarse alignment mode shows the \gls{WFCS} was unable to flatten the fringes. 
		The lower $\mathscr{V}$ pixels on the right side of the pupil vary less between intensity images because a \gls{NPZT} actuator was railed.}\label{fig:ABCD26_28}
    } 
   \label{fig:ABCDflight}
   \end{centering}

\end{figure}

\begin{figure}[htp]
   \centering  
   \includegraphics[width=.9\textwidth]{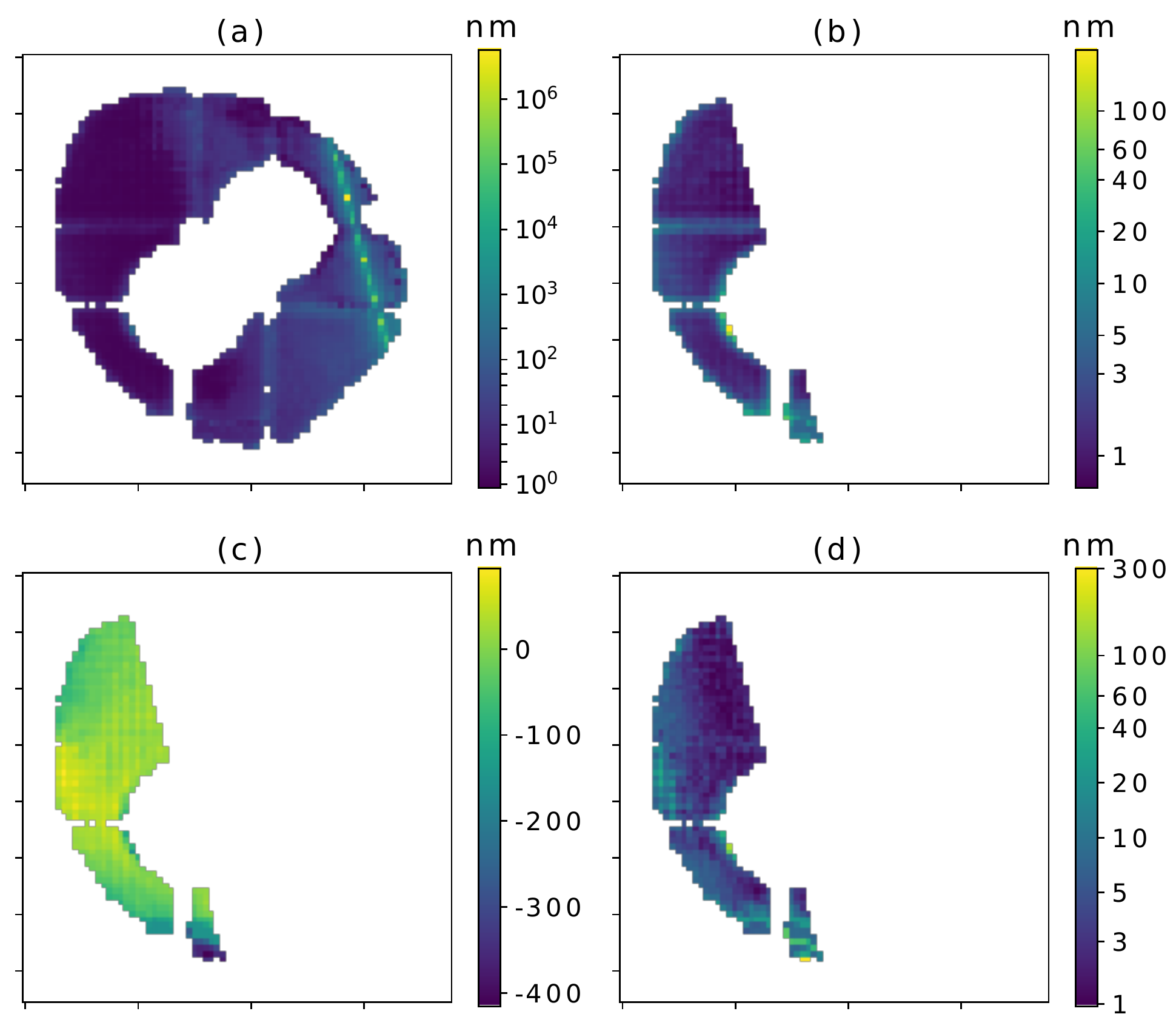}
     \caption{\gls{WFS} pupil phase and error maps showing the phase was poorly measured where the \gls{NPZT} stepsize was smaller than $\pi/4$. 
   Across the region of the pupil where the steps were greater than $\pi/4$ the measured wavefront was flat, with stability of a few nanometers.
     Units are nanometers of wavefront error, calculated by assuming the phase error measured in radians is at 675 nm, the central wavelength of the \gls{VNC}.
     a)  {The mean fitting error across the wavefront sensor. 
     The uncertainty peaks along the axis where the railed \gls{NPZT} actuator held the mirror pinned.}  
     b) {The same as (a), except only showing those pixels where the \gls{NPZT} stepped more than $\pi/4$.  } 
     c) {The mean unwrapped phase after \gls{TTP} terms were subtracted  from each measurement for pixels where the \gls{NPZT} mirror moved more than $\pi/4$.} 
     d) {The \gls{WFS} phase sensing stability, the standard deviation of the measurement-to-measurement phase for pixels where the \gls{NPZT} mirror stepped more than $\pi/4$.}
}\label{fig:lsq_4panel}
\end{figure}

\begin{figure}[htp]
\centering
       \includegraphics[width=0.65\textwidth]{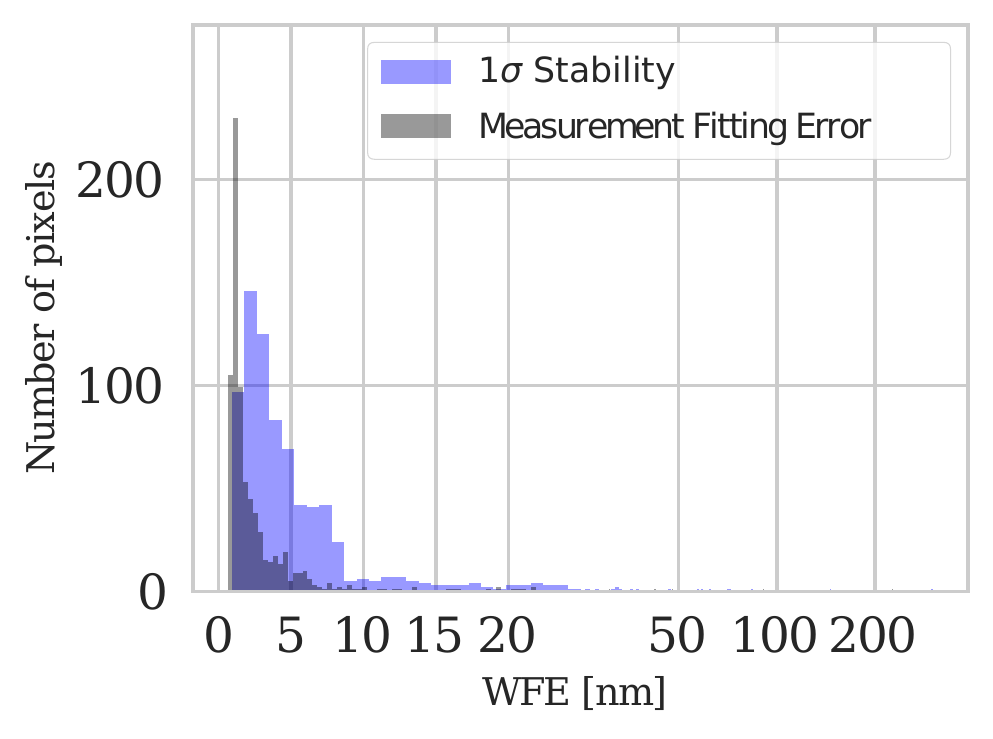}
          \caption{Histograms of the uncertainty in the wavefront sensor pixels where  $\delta$ exceeded $\pi/4$ for the ten high-$\mathscr{V}$ measurements after subtraction of a best-fit plane. 
   The  stability, as measured by  the $1\sigma$ variation between measurements, shows the \gls{DM} and \gls{WFS} were both sufficiently low to reach the expected contrasts. 
   The per-pixel stability variation is significantly larger than the calculated measurement error due to photon noise and fitting errors.}
   \label{fig:wfs_sigma_hist}
\end{figure}

\begin{figure}[htp]
  \centering  
           \includegraphics[width=0.5\textwidth]{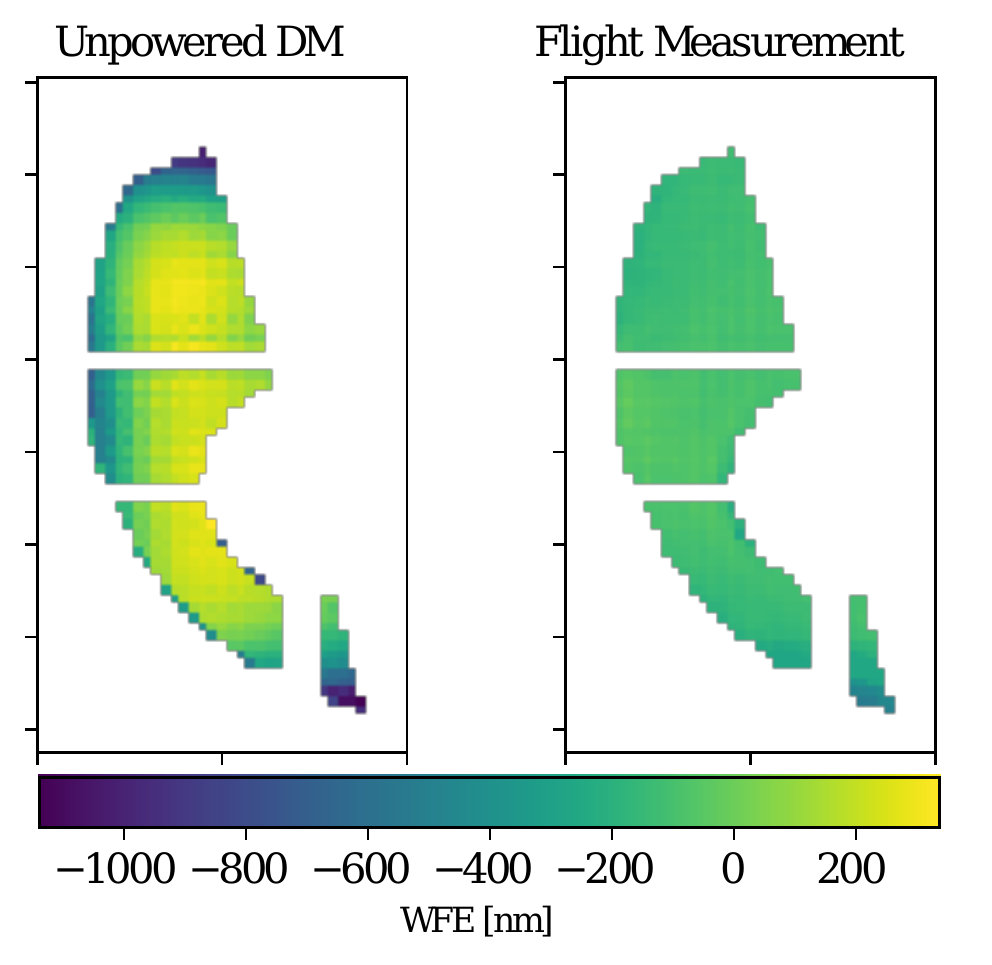}
  \caption{Confirmation of \gls{DM} actuation in flight. 
  Left: Unactuated \gls{WFE} map of the \gls{DM}, digitally masked,  \gls{TTP} subtracted, and interpolated to match the flight wavefront sensor geometry.
  Right: Median wavefront sensor measurement  (Fig. \ref{fig:lsq_4panel}(c)) on the same color scale.
  (Unactuated surface measurement supplied by Boston Micromachines Corporation.) }\label{fig:dm_off_meas}
\end{figure}

\begin{figure}[tp]
   \centering  
   \includegraphics[width=.33\textwidth]{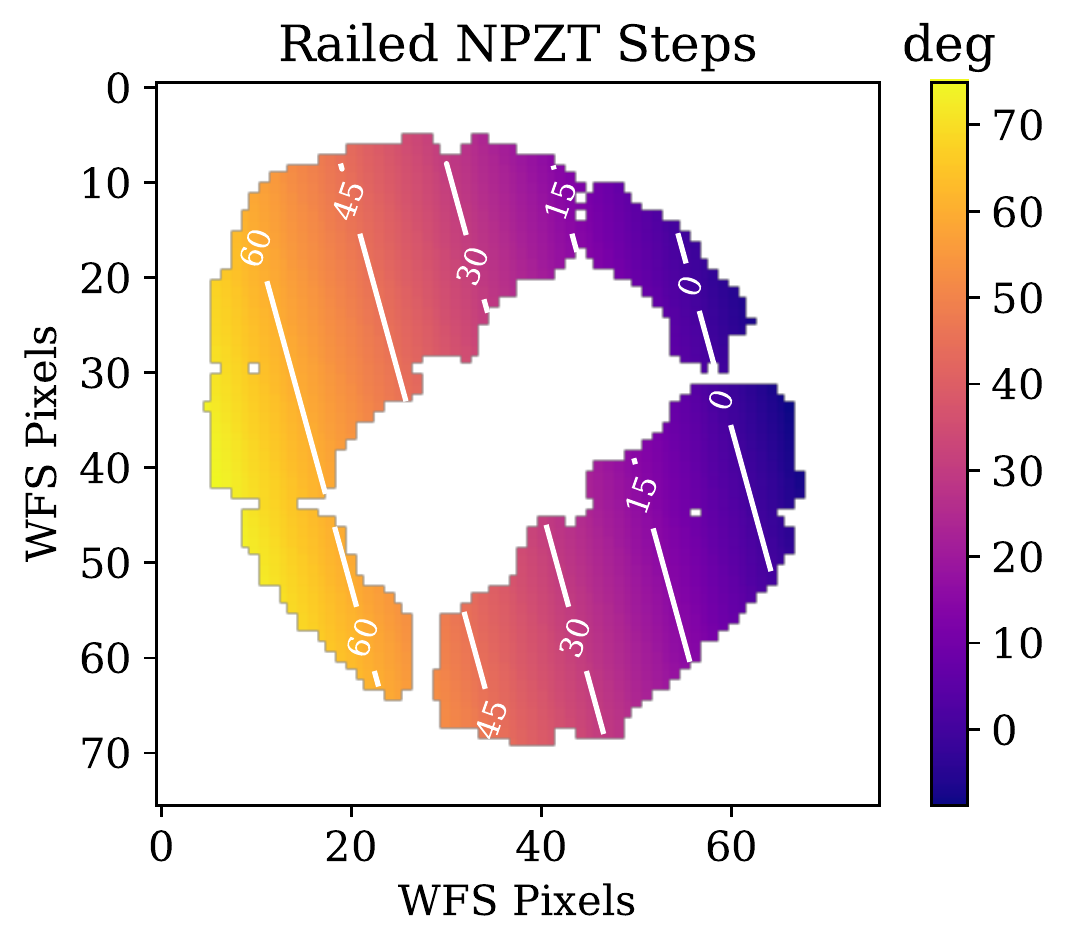}
      \caption{Example of an \gls{NPZT} phase step ($\delta$) map with one actuator driven to its fullest extent, thereby  preventing the mirror to fully step. 
   The resulting phase shift between steps can be seen to pivot about the railed actuator, with the largest shift on the left side of the pupil map. }
   \label{fig:RailedStepMap}
\end{figure}

\begin{figure}[t]
   \caption{Wrapped (left panel) and unwrapped (middle) pupil plane \gls{WFE} measurements from least-squares fitting of four wavefront sensor measurements and corrected \gls{NPZT} positions.
  The uncertainty (right panel) shows $1\sigma$ fitting error including photon noise. 
  }   \label{fig:lsq_phi_maps}
   	\includegraphics[width=\textwidth]{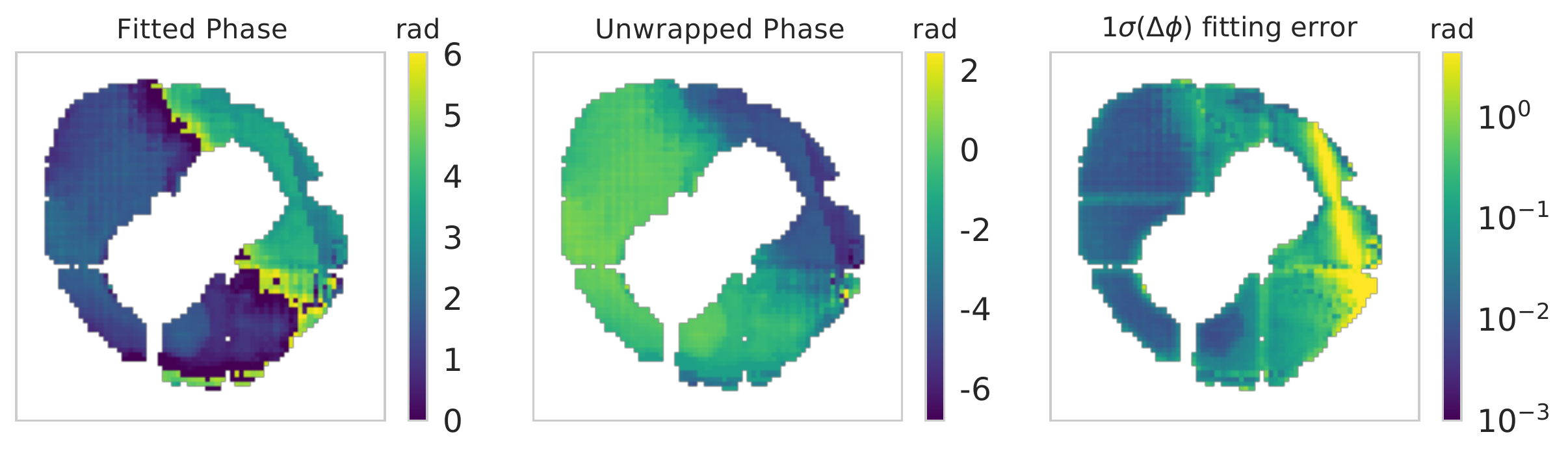}

\end{figure}

\end{document}